\documentclass[prd,onecolumn,nofootinbib,aps,tightenlines,preprintnumbers]{revtex4}

\usepackage{amsmath}
\usepackage{amsfonts}
\usepackage{enumerate}
\usepackage{color}
\usepackage{graphicx}
\usepackage[colorlinks=true,citecolor=blue,urlcolor=blue]{hyperref}

\newcommand{\units}[1]{\ensuremath{\mathrm{#1}}}
\newcommand{\order}[1]{\mathcal{O}(#1)}
\newcommand{\avg}[1]{\left\langle #1 \right\rangle}

\newcommand{\MeV}{\units{MeV}}
\newcommand{\GeV}{\units{GeV}}
\newcommand{\TeV}{\units{TeV}}
\newcommand{\cm}{\units{cm}}

\newcommand{\s}{\units{s}}
\newcommand{\sr}{\units{sr}}
\newcommand{\pb}{\text{pb}}

\graphicspath{ {../Figures/} }

\begin{document}

\preprint{\hbox{PREPRINT UH511-1244-2015}  }

\title{Indirect detection of dark matter using MeV-range gamma-ray telescopes}
\author{Kimberly K. Boddy}
\author{Jason Kumar}
\affiliation{Department of Physics and Astronomy, University of Hawaii,
  Honolulu, Hawaii 96822, USA}

\begin{abstract}
The astrophysics community is considering plans for a variety of gamma-ray telescopes (including ACT, GRIPS, and AdEPT) in the energy range 1--100 MeV, which can fill in the so-called ``MeV gap'' in current sensitivity.
We investigate the utility of such detectors for the study of low-mass dark matter annihilation or decay.
For annihilating (decaying) dark matter with a mass below about 140 MeV (280~MeV) and couplings to first generation quarks, the final states will be dominated by photons or neutral pions, producing striking signals in gamma-ray telescopes.
We determine the sensitivity of future detectors to the kinematically allowed final states.
In particular, we find that planned detectors can improve on current sensitivity to this class of models by up to a few orders of magnitude.
\end{abstract}
\maketitle

\section{Introduction}

The evidence for nonbaryonic, cold dark matter is overwhelming, but there has yet to be any clear indication that dark matter interacts with the Standard Model (SM).
If such interactions exist, it is possible to obtain indirect evidence for dark matter by detecting SM particles that are the products of dark matter annihilations or decays.
The primary difficulty is discriminating these particles from the astrophysical foregrounds, which themselves are not always well understood.
Focusing on sharp or distinct spectral features is a straightforward way to enhance the signal-to-background ratio and boost detector sensitivity.
In particular, a monochromatic gamma-ray line would be a reliable indirect detection signal, since it would be difficult to explain such a signal with conventional astrophysics~\cite{Bergstrom:1997fj}.

Searching for gamma-ray lines as dark matter signatures is a well-studied topic (e.g., see Ref.~\cite{Bringmann:2012ez}).
The most recent searches with gamma-ray satellites%
\footnote{Ground-based Imaging Atmospheric Cherenkov Telescopes are also able to perform indirect-detection searches in dwarf spheroidal galaxies, but their energy thresholds are $\sim 100~\GeV$, above our energy range of interest.}
have been performed by the Fermi-LAT Collaboration~\cite{Abdo:2010nc,Abdo:2010dk,Ackermann:2012qk}.
The upcoming space-based telescope GAMMA-400~\cite{Galper:2014pua} is set to launch in 2018 and will have an energy range that overlaps with Fermi and extends up to $3~\TeV$.
There is, however, a significant ``MeV gap''~\cite{Greiner:2011ih} in gamma-ray detector sensitivity in the $0.1$--$100~\MeV$ range.
There have been proposals to address this gap with future experiments, such as the Advanced Compton Telescope (ACT)~\cite{Boggs:2006mh}, the Advanced Pair Telescope (APT)~\cite{Buckley:2008fk}, the Gamma-Ray Imaging, Polarimetry and Spectroscopy (GRIPS) detector~\cite{Greiner:2011ih}, the Advanced Energetic Pair Telescope (AdEPT)~\cite{Hunter:2013wla}, the Pair-Production Gamma-Ray Unit (PANGU)~\cite{Wu:2014tya}, and the Compton Spectrometer and Imager (COSI)~\cite{Kierans:2014lr}.
Also, the ASTROGAM Collaboration\footnote{\url{http://astrogam.iaps.inaf.it/index.html}.} is currently working on the research and development of a space-based mission for an MeV instrument.
Given these efforts by the astrophysics community, it is especially interesting to consider models of low-mass dark matter, which can produce distinctive photon signatures in this energy range, particularly since the latest direct-detection experiments (such as LUX~\cite{Akerib:2013tjd}, Super-CDMS~\cite{Agnese:2014aze}, and CDMSlite~\cite{Agnese:2013jaa}) lose sensitivity for $m_X \lesssim ~ \GeV$, making indirect searches all the more important.

There is a class of dark matter models that will generally produce distinctive photon signatures in the 1--100$~\MeV$ range: low-mass ($\sqrt{s} \lesssim 280~\MeV$) annihilating/decaying dark matter that couples significantly to first-generation quarks.
For such models, the annihilation/decay products are tightly constrained by kinematics; the only available nonleptonic two-body final states are $\gamma \gamma$, $\gamma \pi^0$, and $\pi^0 \pi^0$.
These channels will dominate in most of the parameter space, and all of these channels produce sharp features, such as lines or boxes, in the photon spectrum.
Our focus will be on the constraints one may place on these channels from current data and from future telescopes.

While the annihilation/decay of dark matter to $\gamma \gamma$ is well studied, the final states $\gamma \pi^0$ and $\pi^0 \pi^0$ are less so.
Similar studies have investigated monoenergetic photons produced in the processes $XX \rightarrow \gamma \phi$ and $X \rightarrow \gamma \phi$, with $h$ or $Z$ playing the role of $\phi$~\cite{Bergstrom:1997fj,Goodman:2010qn,Rajaraman:2012ix,Bergstrom:2012vd}.
Additionally, the cascade process $XX \rightarrow \phi\phi \rightarrow \gamma\gamma\gamma\gamma$ produces a box-shaped spectrum from $\phi$ decay, and Fermi-LAT data can constrain such channels for dark matter masses above $5~\GeV$~\cite{Ibarra:2012dw}.
In our case, however, we consider $\pi^0$ in the role of $\phi$, at a much different energy range.
There have also been models of dark matter decay that give monoenergetic photons in our energy range of interest~\cite{Yuksel:2007xh,Yuksel:2007dr,Bell:2010fk}.

The plan of this paper is as follows.
In Sec.~\ref{sec:model} we introduce the class of low-mass dark matter models that give rise to sharp gamma-ray spectral features, and in Sec.~\ref{sec:cosmology} we briefly discuss the cosmological bounds on these models.
We then go into details in Sec.~\ref{sec:flux} on the astrophysical signatures for the indirect detection of dark matter.
The specific systems we focus on are the gamma-ray diffuse background in Sec.~\ref{sec:diffuse} and dwarf spheroidal galaxies in Sec.~\ref{sec:dwarf}.
We comment on possible collider constraints in Sec.~\ref{sec:collider} and conclude in Sec.~\ref{sec:conclusions}.

\section{Dark Matter Interactions}
\label{sec:model}

We consider low-mass dark matter to be a particle $X$ that couples directly to first-generation SM quarks only.
Generally, we can consider either dark matter annihilation or decay to SM final states.
Since the dark matter initial state is assumed to be neutral under all unbroken gauge symmetries of the SM, the only remaining relevant symmetries to consider are $C$, $P$, and $T$.
The only possible two-body final states that are kinematically accessible are as follows.
\begin{enumerate}[(i)]
\item $\gamma \gamma$: Accessible at all energies. The final state is $C$-even.
\item $\gamma \pi^0$: Accessible for $\sqrt{s} > m_{\pi^0}$. The final state is $C$-odd.
\item $\pi^0 \pi^0$: Accessible for $\sqrt{s} \geq 2m_{\pi^0}$. The final state is $C$-even.
\item $\pi^+ \pi^-$: Accessible for $\sqrt{s} \geq 2m_{\pi^\pm}$. The final state is $C$-even or $C$-odd.
\item $\bar \ell \ell$ ($\ell = e, \mu, \nu$): Accessible for $\sqrt{s} \geq 2m_{\ell}$. The final state is either $C$-odd or is weak suppressed.
\end{enumerate}
Note that if $X$ decays rather than annihilates to SM particles, then these final states are only allowed if $X$ is a boson.
If we assume that weak interactions, which are suppressed by a factor $s G_F$, are negligible, then the only leptonic states which can be produced are the $C$-odd $e^+ e^-$ and $\mu^+ \mu^-$ final states via an intermediate off-shell photon that couples to a quark loop.
But the annihilation/decay rate in this channel will be suppressed by a factor $\alpha^2$, so it will be dominant only if no other channel can be produced.
Since the photon spectrum arising from this channel is not as distinctive as the others, we will ignore it from here on.
Three-body final states are also accessible if additional photons or neutrinos are emitted; however, these processes come with additional suppression factors of $\alpha$ or $s G_F$, respectively, and can thus also be ignored unless all of the channels above are heavily suppressed.
It is worth mentioning that final-state radiation (FSR) and virtual internal bremsstrahlung (VIB) are capable of producing somewhat distinctive features in the photon spectrum.
FSR yields a continuous $1/E$ spectrum~\cite{Birkedal:2005ep}, distinct from the astrophysical foregrounds that have a spectral index around $-2$ (see Sec.~\ref{sec:diffuse-future}).
It may be possible to observe the upper edge of such a spectrum, determined by $m_X$ for annihilation or $m_X/2$ for decay~\cite{Bell:2008vx}.
Similarly, VIB can produce a broad bump near $E_\gamma \lesssim m_X$ (for annihilation), which mimics a line in a detector with poor energy resolution, but the details of the spectrum tend to be more model dependent~\cite{Bringmann:2007nk}.

For $\sqrt{s} \geq 2m_{\pi^\pm}$ it is kinematically possible to produce the $\pi^+ \pi^-$ final state.
As with charged leptons, FSR can produce a sharp edge in the photon spectrum; however, this process could potentially compete with the $\gamma\pi^0$ or $\pi^0\pi^0$ channel, diminishing the prospect of observation.
Moreover, even though these states with one or two neutral pions could be produced with non-negligible branching fractions, their photon spectrum becomes very broad if $\sqrt{s}$ is significantly greater than $2m_{\pi^0}$ and thus less interesting from the point of view of detectability at future experiments.
Additionally, one would expect the branching fraction to the $\gamma \gamma$ final state to be suppressed by $\sim \alpha^2$ relative to the $\pi^+ \pi^-$ state.
We thus see that this class of models produces the most interesting photon spectrum, from the point of view of detectability, for the range $\sqrt{s} \lesssim 2m_{\pi^\pm}$.
We limit our study to this mass range.

If we assume that weak interactions are negligible, then the only potential source of $C$ violation is via the coupling of dark matter to the first-generation quarks.
If we insist that this coupling is instead $C$ invariant, then one may classify allowed final states by the $C$ quantum number of the initial state.
For example, in a $C$-invariant effective field theory, a $C$-odd initial dark matter state can only produce the $\gamma \pi^0$ or $\bar \ell \ell$ two-body final states; if the $\gamma \pi^0$ state is kinematically allowed, then it will dominate.
An example of this type of fundamental interaction (if $X$ is a Dirac fermion) would be the interaction $(1/M^2)(\bar X \gamma^\mu X)(\bar q \gamma_\mu q)$, where $q=u,d$.
In this case, the $\bar X X$ initial state with a nonvanishing matrix element is $C$ odd.
In the low-energy effective field theory involving only $X$, $\gamma$, and $\pi^0$ (excluding leptons), the lowest-dimension Lagrangian interaction one can write that is Lorentz and $C$ invariant and linear in $\bar X \gamma^\mu X$ is
\begin{equation}
  \mathcal{L} \sim \frac{1}{\Lambda^3} (\bar{X} \gamma^\mu X)
  F^{\nu\rho} (\partial^\sigma \pi^0) \epsilon_{\mu\nu\rho\sigma} \ .
\label{eq:VV_eff_op}
\end{equation}
The coefficient $\Lambda$ can be related to the one-loop diagram mediating the $\pi^0 \rightarrow \gamma \gamma$ decay interaction (by replacing one photon with the dark matter vector current), which is determined by the chiral anomaly.%
\footnote{The $X/\pi^0/\gamma$ coupling is not related via isospin to any potential coupling of dark matter to $\pi^\pm$, because electromagnetic interactions violate isospin near-maximally.}
We would then expect $1/\Lambda^3 \sim (e/16\pi^2)(1/M^2 f_\pi)$, where $f_\pi$ is the pion decay constant.

In general, however, the dark matter--SM quark-level interaction need not preserve $C$.
As such, dark matter annihilations (decays) are able to produce all three final states of interest---provided they are kinematically accessible---with branching ratios $\mathcal{A}_\pi$, $\mathcal{A}_{\gamma\pi}$, and $\mathcal{A}_\gamma$ ($\mathcal{D}_\pi$, $\mathcal{D}_{\gamma\pi}$, and $\mathcal{D}_\gamma$), respectively.
These branching ratios depend on the specific UV model, but we leave them here as parameters to keep our analysis general.

\subsection{Photon spectra}

The prompt photons simply have $\delta$-function spectra in the dark matter center-of-mass frame:
\begin{equation}
  \frac{dN_\gamma}{dE} = \delta(E - E_0) \ ,
  \label{eq:line-spectrum}
\end{equation}
for each photon produced with an energy $E_0$.
Any $\pi^0$ subsequently decays to two secondary photons with a branching ratio of $\sim 99\%$~\cite{Agashe:2014kda}.
The decay is isotropic in the $\pi^0$ rest frame, and boosting to the dark matter center-of-mass frame results in a box-shaped photon spectrum~\cite{Stecker:1971qtb}
\begin{equation}
  \frac{dN_\gamma}{dE}
  = \frac{2}{\Delta E} \left[\Theta(E-E_-)-\Theta(E - E_+)\right] \ ,
  \label{eq:box-spectrum}
\end{equation}
where $E_\pm$ are the kinematic edges and $\Delta E \equiv E_+ - E_-$ is the box width.
Thus, the annihilation/decay processes yielding prompt photons or neutral pions produce gamma spectra with sharp features.
We summarize the kinematics below.
\begin{enumerate}[(i)]
\item $\pi^0 \pi^0$:
  The photon spectrum is box shaped, given by twice that in Eq.~\eqref{eq:box-spectrum}, with kinematic edges and box width
  \begin{equation}
    E_\pm = \frac{\sqrt{s}}{4}
    \left(1\pm\sqrt{1-\frac{4m_{\pi^0}^2}{s}}\right) \ , \qquad
    \Delta E = \sqrt{\frac{s}{4}-m_{\pi^0}^2} \ .
  \end{equation}
\item $\gamma \pi^0$:
  The prompt photon produces a line distribution, given by Eq.~\eqref{eq:line-spectrum}, with energy
  \begin{equation}
    E_0 = \frac{\sqrt{s}}{2} \left(1-\frac{m_{\pi^0}^2}{s}\right) \ .
  \end{equation}
  The spectrum from the pion decay is given by Eq.~\eqref{eq:box-spectrum}, with kinematic edges and box width
\begin{equation}
  E_\pm = \frac{\sqrt{s}}{4} \left[
    \left(1+\frac{m_{\pi^0}^2}{s}\right) \pm
    \left(1-\frac{m_{\pi^0}^2}{s}\right)\right] \ , \qquad
  \Delta E = \frac{\sqrt{s}}{2} \left(1-\frac{m_{\pi^0}^2}{s} \right) \ .
\end{equation}
\item $\gamma\gamma$:
  The photon spectrum is a line, given by twice that in \eqref{eq:line-spectrum}, at the energy
  \begin{equation}
    E_0 = \frac{\sqrt{s}}{2} \ .
  \end{equation}
\end{enumerate}
If the box spectrum is very narrow, a detector will not be able to resolve the box shape and will instead observe a signal that is indistinguishable from a line.
Since the box width is larger for a more highly boosted pion, a pion produced nearly at rest ($\sqrt{s}/2 \approx m_{\pi^0}$ for the $\pi^0 \pi^0$ channel and $\sqrt{s} \approx m_{\pi^0}$ for the $\gamma\pi^0$ channel) produces two photons with energies close to $m_{\pi^0}/2$.
At the upper end of the kinematic range we consider ($\sqrt{s} = 2m_{\pi^\pm}$), the width of the box spectrum for the $\gamma \pi^0$ and $\pi^0 \pi^0$ channels is $\sim 106.9~\MeV$ and $\sim 35.5~\MeV$, respectively.

\section{Cosmological Bounds}
\label{sec:cosmology}

Searches for particle dark matter are typically performed by looking for signals from production at colliders, direct detection, or indirect detection.
Collider constraints are highly model dependent, and later in Sec.~\ref{sec:collider} we will briefly discuss bounds in the context of the example model presented in Sec.~\ref{sec:model}.
But since we are interested in the broader class of models, we do not consider collider constraints in our overall analysis.
Current direct-detection experiments are sensitive to dark matter masses above $\sim 1~\GeV$, so their results do not apply to dark matter at much lower masses.
However, the annihilation/decay of light dark matter can be constrained by its effect on the photon spectrum observed by indirect-detection experiments, which is the main focus this paper.

Another constraint arises from the effects of dark matter annihilation/decay on cosmological history.
Before investigating the indirect detection of dark matter, we first discuss these limits from the early Universe.
Dark matter annihilations and decays can inject energy into the primordial plasma, altering the predictions of big bang nucleosynthesis (BBN)~\cite{Cyburt:2002uv,Kawasaki:2004qu,Ellis:2005ii,Jedamzik:2006xz,Hisano:2009rc,Henning:2012rm} and the spectrum of the cosmic microwave background (CMB)~\cite{Padmanabhan:2005es,Mapelli:2006ej,Galli:2009zc,Slatyer:2009yq,Cirelli:2009bb,Galli:2011rz,Finkbeiner:2011dx,Slatyer:2012yq,Madhavacheril:2013cna,Cline:2013fm,Lopez-Honorez:2013cua,Diamanti:2013bia}.
For dark matter masses $\order{100~\MeV}$, CMB limits for $s$-wave annihilation are more stringent than those from BBN~\cite{Henning:2012rm}.
If we consider the latest Planck limits~\cite{Ade:2015xua}, the upper bound on the thermalized annihilation cross section is
\begin{equation}
  \avg{\sigma v} \lesssim 3\times 10^{-28}
  \left(\frac{m_X}{\GeV}\right)~\cm^3/\s \ .
  \label{eq:CMBbound}
\end{equation}
Here we assume that dark matter annihilation produces only photons, since the neutral pions that are produced rapidly decay into photons on a time scale less than the Hubble time scale during recombination.
This limit is about an order of magnitude stronger than the previous WMAP9 limit~\cite{Madhavacheril:2013cna}.
If the annihilation process is instead $p$-wave, the CMB limits weaken~\cite{Diamanti:2013bia} and present-day limits from the diffuse gamma-ray background can dominate~\cite{Essig:2013goa}.
In this case, future gamma-ray telescopes providing better measurements of the diffuse background can also better constrain dark matter $p$-wave annihilations.
The more challenging question is whether or not future telescopes will have sensitivities to reach the strong bound in Eq.~\eqref{eq:CMBbound}, so here we focus on the simpler analysis of $s$-wave annihilation.
For decaying dark matter, we use the numerical tools presented in Ref.~\cite{Slatyer:2012yq} to analyze the case of dark matter decaying to monoenergetic photons, and the resulting lower bound on the lifetime from Planck is $\tau \gtrsim 10^{23}$--$10^{24}~\s$.
The annihilation and decay bounds are displayed as black lines in the left and right panels, respectively, of Fig.~\ref{fig:constraints}.

Note that the most stringent current constraints on dark matter annihilation come from its effect on the CMB, while CMB constraints on dark matter decay are less stringent than those arising from the observed diffuse gamma-ray background, as we discuss in Sec.~\ref{sec:diffuse-existing}.
CMB constraints are much more severe for the case of dark matter annihilation, because the annihilation rate is proportional to the dark matter density squared, while the decay rate is proportional to just a single power of the density; the annihilation rate thus has a large enhancement in the early Universe relative to the current epoch.

\section{Gamma-Ray Flux}
\label{sec:flux}

The differential flux of gamma rays from annihilating dark matter is
\begin{equation}
  \frac{d^2 \Phi}{dE\, d\Omega}
  = \frac{J}{4\pi} \frac{\avg{\sigma v}}{2m_X^2} \sum_i \mathcal{A}_i
  \int \frac{dN_\gamma^i(E')}{dE'} R_\epsilon (E-E') \, dE' \ ,
  \label{eq:diff2flux}
\end{equation}
where $m_X$ is the dark matter mass, $\avg{\sigma v}$ is the velocity-averaged annihilation cross section, and $R_\epsilon$ is the smearing function due to the  limited detector energy resolution $\sigma_E = \epsilon E$.
Here we assume that dark matter is its own antiparticle; if it is not, then the expression for the flux is rescaled by a factor of $1/2$.
We estimate the detector smearing function with a Gaussian, which must be convolved with the true spectrum.
For a gamma ray with true energy $E'$, the probability that it will be reconstructed with an energy $E$ is~\cite{Bringmann:2008kj,Perelstein:2010at}
\begin{equation}
  R_\epsilon (E-E') \approx \frac{1}{\sqrt{2\pi}\epsilon E'}
  \exp\left(-\frac{(E-E')^2}{2\epsilon^2 E^{\prime 2}}\right) \ .
  \label{eq:resp_func}
\end{equation}
As a result, an injected monochromatic line distribution will appear at the detector as a Gaussian distribution, centered at the line energy $E_0$, with a half-width given by $\epsilon E_0$.
The kinematic edges of the pion decay box-shaped spectrum will be smeared out as well.
The sum in Eq.~\eqref{eq:diff2flux} runs over all dark matter annihilation processes $i$ with branching ratios $\mathcal{A}_i$ and spectra $dN_\gamma^i / dE$.
The total gamma-ray spectrum has contributions from continuum spectra and from monochromatic lines, and we are only concerned with spectra that are isotropic.

All of the terms in Eq.~\eqref{eq:diff2flux} depend only on the particle properties of the dark matter, except for the astrophysical $J$-factor
\begin{equation}
  J(\psi) \equiv \int_\textrm{LOS} \rho(r(l,\psi))^2 \, dl \ ,
  \label{eq:J-factor}
\end{equation}
where the integral is taken over the line of sight, and $\rho(r)$ is the dark matter energy density distribution.
The coordinate $r$ is centered on the dark matter halo of interest, which is a distance $D$ from the sun.
The quantity $\psi$ is the angle between the line of sight and the line connecting the sun with the center of the halo.
Thus, we have $r=(D^2 -2Dl \cos\psi + l^2)^{1/2}$, and the line-of-sight integral is from 0 to $l_\textrm{max}=(R_\textrm{halo}^2 - \sin^2\psi D^2)^{1/2}+D \cos\psi$~\cite{Yuksel:2007ac}.
To find the total dark matter flux from a particular region of the sky, we need the $J$-factor integrated over that region:
\begin{equation}
  \mathcal{J} \equiv \int d\Omega_\psi \, J(\psi) \ .
\end{equation}

The total number of photons with energies between $E_1$ and $E_2$ detected from a source with an integrated $J$-factor given by $\mathcal{J}$, observed over a period of time $T_\textrm{obs}$, is
\begin{equation}
  N_\gamma = T_\textrm{obs} A_\textrm{eff} \Phi
  = T_\textrm{obs} A_\textrm{eff} \frac{\mathcal{J}}{4\pi}
  \frac{\avg{\sigma v}}{2m_X^2} \sum_i \mathcal{A}_i
  \int_{E_1}^{E_2} \frac{dN_\gamma^i(E')}{dE'} R_\epsilon (E-E') \, dE' \ ,
  \label{eq:Nphotons}
\end{equation}
where $A_\textrm{eff}$ is the effective area of the detector.
The analogous expressions for decaying dark matter are obtained by making the substitutions $\avg{\sigma v}/2m_X \to \Gamma$ and $\mathcal{A}_i \to \mathcal{D}_i$ in Eqs.~\eqref{eq:diff2flux} and \eqref{eq:Nphotons}, and $\rho^2 \to \rho$ in Eq.~\eqref{eq:J-factor}, where $\Gamma$ is the decay width and the sum runs over decay processes $i$ with branching ratios $\mathcal{D}_i$.

There are a few ideal systems to search for an indirect-detection signal from dark matter: dwarf spheroidal galaxies, the diffuse gamma-ray background, and the Galactic center (GC).
We do not focus on regions near the GC in this work, due to the large astrophysical uncertainties involved.
For example, the uncertainty in the spectrum arising from astrophysical backgrounds is very difficult to quantify.
In addition, the central halo profile is uncertain due to effects from a supermassive black hole~\cite{Gondolo:1999ef,Merritt:2002vj,Bertone:2005hw,Bertone:2005xv}, from interactions with baryons~\cite{Gnedin:2003rj,Tonini:2006uqa}, and from baryonic infall~\cite{Mo:1997vb,Gnedin:2004cx}.

In the following sections, we describe the properties of the other two systems and their potential for producing signals from dark matter.
We treat the three dark matter annihilation/decay channels of interest separately, assuming in each case that the branching ratio is 100\%.
For each of these channels, we determine limits from the diffuse background for previous and existing experiments (COMPTEL, EGRET, and Fermi), and we give projected limits from both the diffuse background and dwarfs for future experiments (ACT, GRIPS, AdEPT, and GAMMA-400).
The general detector properties we use are summarized in Table~\ref{tab:experiments}.
We use more conservative estimates of energy and angular resolution (in the most relevant energy regions) provided by the experiments.
Actual detectors have complex response functions (e.g., $\sigma_E$ is a more complicated function of energy, and $A_\textrm{eff}$ depends on energy and detected angle of incidence) and are typically optimized in the central region of their energy window.
Given that we are mainly interested in the potential of future experiments rather than setting the most precise constraints on existing experiments, we use these baseline numbers for a more straightforward analysis.

\begin{table}[t]
  \centering
  \begin{tabular}{cccccc}
    Detector & Source & Energy Range [MeV] & $\epsilon$
    & PSF & $A_\textrm{eff}$ [$\cm^2$] \\
    \hline
    \hline
    ACT & \cite{Boggs:2006mh}
    & 0.2--10 & 1\% & $1^\circ$ & 1000 \\
    GRIPS & \cite{Greiner:2011ih}
    & 0.2--80 & 3\% & $1.5^\circ$ & 200 \\
    AdEPT & \cite{Hunter:2013wla}
    & 5--200 & 15\% & $0.5^\circ$ & 600 \\
    COMPTEL & \cite{Weidenspointner:1999thesis,Kappadath:1998thesis}
    & 0.8--30 & 2\% & $2^\circ$ & 50  \\
    EGRET & \cite{Thompson:1993lr}
    & $30~\MeV$--$10~\GeV$ & 12.5\% & $2.8^\circ$ & 1000 \\
    Fermi-LAT & \cite{Atwood:2009ez}
    & $20~\MeV$--$300~\GeV$ & 7.5\% & $2^\circ$ & 4000 \\
    GAMMA-400 & \cite{Galper:2014pua}
    & $100~\MeV$--$3~\TeV$ & 12\% & $2^\circ$ & 3000
  \end{tabular}
  \caption{Experimental parameters used to determine indirect-detection bounds for dark matter annihilation and decay.
  The PSF and $A_\textrm{eff}$ values are not needed in the analysis for COMPTEL, EGRET, and Fermi, but they are included for comparison.}
  \label{tab:experiments}
\end{table}

\section{Indirect Detection: Diffuse Gamma-Ray Background}
\label{sec:diffuse}

One way to search for dark matter is observing the diffuse gamma-ray background.
There are two components to the diffuse background: Galactic and extragalactic.
We expect the diffuse background to comprise signals from both dark matter and astrophysical processes, but the astrophysical contribution cannot be estimated entirely from data: one must instead make some assumptions about an underlying astrophysics model.
Additionally, there will be enhancements for Galactic and extragalactic dark matter annihilation due to substructure~\cite{Ullio:2002pj,Strigari:2006rd,Zavala:2009zr}; however, different models can yield a wide range of results~\cite{Abdo:2010dk}, so we ignore substructure enhancements and obtain more conservative bounds on dark matter annihilation.

The Galactic contribution from dark matter is simply given by Eq.~\eqref{eq:diff2flux}.
We use publicly provided calculations of $J(\psi)$~\cite{Cirelli:2010xx} with an Nevarro-Frenk-White profile; far from the GC, the $J$-factor becomes less sensitive to the details of the central profile, and the integrated $J$-factors for various profiles differ by $\order{1}$ factors at high Galactic latitudes~\cite{Cirelli:2010xx}.
For the extragalactic background, we assume the dark matter contribution is isotropic.
Photons detected with energy $E_0$ were emitted at redshift $z$ with energy $E(z)=(1+z)E_0$.
Although redshifting has the effect of smearing out an otherwise sharp signal from a monoenergetic photon or from neutral pion decay, the extragalactic flux is subdominant to the Galactic flux, particularly in the case of annihilation.
Photons emitted at redshift $z \ll 1$ are relevant, because they provide a small enhancement on top of the Galactic signal, within the energy resolution of the detector.

\subsection{Limits from existing data}
\label{sec:diffuse-existing}

For the energy range of interest, COMPTEL, EGRET, and Fermi-LAT give the most relevant data of the differential flux for the Galactic and extragalactic spectrum%
\footnote{Although COMPTEL presents results for the extragalactic diffuse spectrum, the given measurements do not have the Galactic spectrum or point sources removed due to large systematic uncertainties~\cite{Weidenspointner:1999thesis}.}
at high Galactic latitudes $b$.
COMPTEL~\cite{Weidenspointner:1999thesis} observed the region $|b|>30^\circ$, EGRET~\cite{Strong:2004de} observed $20^\circ<|b|<60^\circ$ (see their Fig.~8, region~E), and Fermi~\cite{Ackermann:2014usa} observed $|b|>20^\circ$ (see their Fig.~4).
Note that although Fermi is capable of detecting photons down to energies of $20~\MeV$, the data for the diffuse analysis only goes down to energies of $100~\MeV$ in order to sufficiently reduce the cosmic-ray background produced in the Earth's atmosphere.

\begin{figure}[t]
  \includegraphics[scale=0.6]{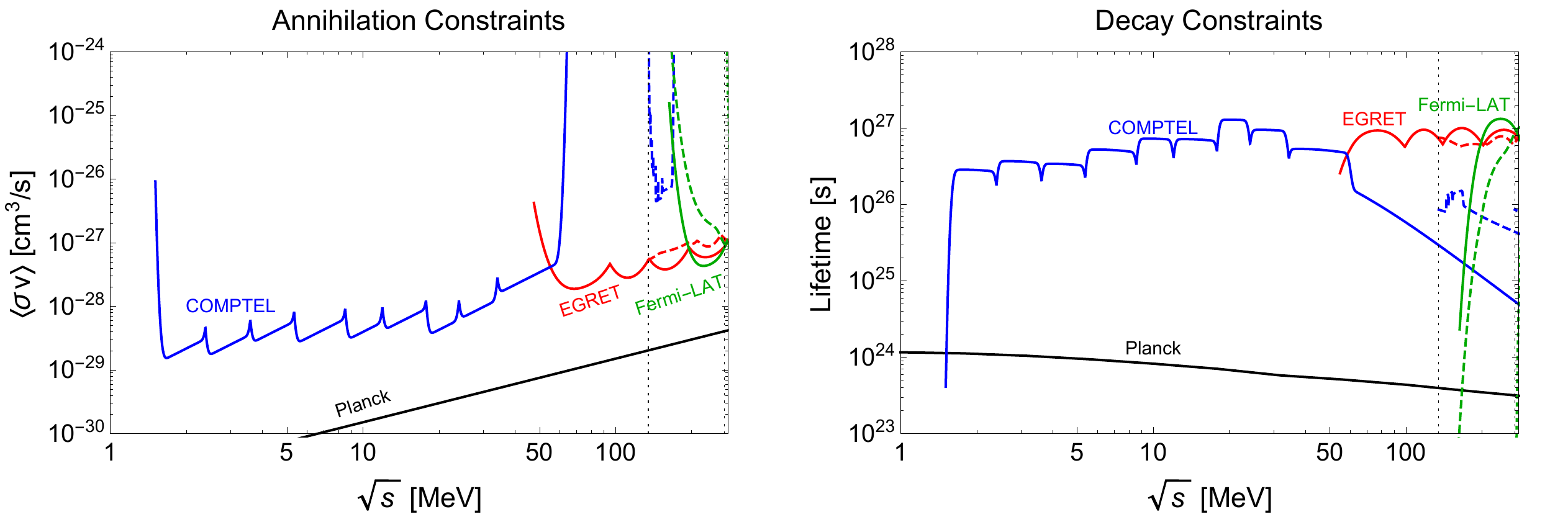} \\
  \includegraphics[scale=0.6]{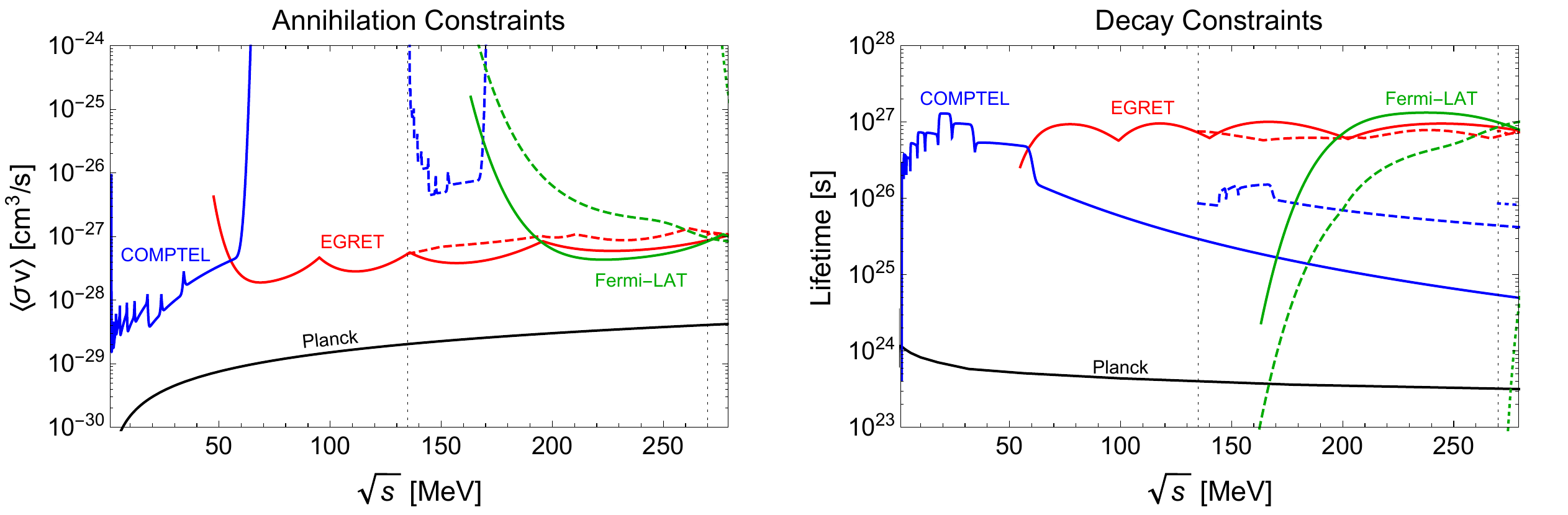}
  \caption{CMB and diffuse constraints on dark matter annihilation (left) and decay (right).
  The top and bottom sets of panels display the same information, plotted on a log and linear $x$ axis, respectively.
  The strongest CMB bound comes from Planck, shown in black for annihilation~\cite{Ade:2015xua} and decay~\cite{Slatyer:2012yq}.
  Conservative bounds from COMPTEL, EGRET, and Fermi for diffuse emission are shown in blue, red, and green, respectively, for the channels $\gamma\gamma$ (solid), $\gamma\pi^0$ (dashed), and $\pi^0 \pi^0$ (dotted).
  Optimistic bounds, described in Sec.~\ref{sec:diffuse-future}, can be obtained by improving the COMPTEL, EGRET, and Fermi bounds by a factor of $\sim$3--5, 9, and 6, respectively.
  Vertical dotted lines show the kinematic thresholds at $m_{\pi^0}$ and $2m_{\pi^0}$, and $\sqrt{s}$ is plotted up to $m_{\pi^\pm}$.}
  \label{fig:constraints}
\end{figure}

To determine a conservative limit on the lifetime (cross section) for decaying (annihilating) dark matter, we follow the procedure from Ref.~\cite{Essig:2013goa}: we require that the number of events expected from a dark matter signal in each energy bin does not exceed the observed number of counts by $2\sigma$.
The resulting limits are shown in Fig.~\ref{fig:constraints} for each of the three channels of interest: $\gamma\gamma$ (solid), $\gamma\pi^0$ (dashed), and $\pi^0 \pi^0$ (dotted).
Prompt photons will typically fall within one or two energy bins (depending on the effects of the energy resolution and the bin placement relative to the photon energy).
Photons arising from $\pi^0$ decay will fall roughly uniformly within the the energy range bounded by $E_-$ and $E_+$; the most important energy bin for these photons is the highest energy bin within this range, since the observed flux falls as a function of energy.

The effect of the discrete binning is clear, particularly for the $\gamma\gamma$ channel, and the rounded shape for each bin is due to the smearing effects of the detector energy resolution%
\footnote{The sizes of the bins chosen by each experiment are larger than the minimum sizes determined by the energy resolution.}.
For the $\gamma\pi^0$ channel, the projected sensitivity of COMPTEL is dominated by the monoenergetic photon and is best at $\sqrt{s} \sim 160~\MeV$.
Just above threshold in either the $\gamma \gamma$ or $\gamma \pi^0$ channels, COMPTEL's sensitivity drops rapidly because any monoenergetic photon is below COMPTEL's energy range, while any photons arising from pion decay are nearly monoenergetic and are above COMPTEL's energy range.
For either the $\gamma\pi^0$ or $\pi^0 \pi^0$ channel, COMPTEL's sensitivity to the pion is very weak, since the box feature in the spectrum must be very wide to overlap COMPTEL's energy range.
For the $\gamma\gamma$ channel at large $\sqrt{s}$, COMPTEL's sensitivity is dominated by the extragalactic diffuse component, since the gamma rays at low redshift have energies above COMPTEL's energy range.
The limits from EGRET and Fermi are much simpler, since $m_{\pi^0}$ lies within their energy ranges.

It is worth noting that a conservative GC analysis for COMPTEL and EGRET~\cite{Mack:2008wu} is able to set limits on dark matter annihilation to $\gamma\gamma$ that are comparable to the high Galactic latitude bounds in Fig.~\ref{fig:constraints}.
There is another GC analysis for EGRET~\cite{Pullen:2006sy} that constrains the annihilation cross section to $\gamma \gamma$ at the level $\langle \sigma v \rangle \lesssim 10^{-30}~\cm^2$ for $\sqrt{s} = 100~\MeV$.
However, this analysis presupposes that the photon spectrum arising from astrophysical background is smooth and can be determined with subleading systematic uncertainties by fitting to data; the allowed dark matter contribution arises only from a $2\sigma$ downward statistical fluctuation in the background in any bin.
This analysis is thus not as conservative as either the conservative analysis approach used above (and used by Ref.~\cite{Mack:2008wu}) or the optimistic analysis described in Sec.~\ref{sec:diffuse-future}, in which systematic uncertainties dominate.

\subsection{Limits from future experiments}
\label{sec:diffuse-future}

For the existing experiments, we used their energy binning for their data.
In analyzing future experiments, we use an optimized binning strategy in an effort to obtain the best sensitivity.
Practically, doing so avoids the discreteness (the small dips) of the curves in Fig.~\ref{fig:constraints}.
The relevant energy bins are the ones encompassing the monoenergetic photon and/or the box feature in the photon spectrum arising from pion decay.
The bin for the prompt photon is one centered at its true energy $E_0$ and has a width $2\epsilon E_0$ (i.e., the bin detects 68\% of the line signal).
For the box spectrum, the energy bin that gives the best constraint is the one at the upper edge of the box, so we define a bin with an upper edge at $E_+$.
In either case, if the signal leaks beyond the lower or upper detector threshold, the bin edge is set to be at the threshold.
For either the $\gamma\gamma$ or $\pi^0\pi^0$ channel, the only energy bin to consider is the one corresponding to the line or upper box edge, respectively.
For the $\gamma\pi^0$ channel, the more constraining bin (if it is inside the detector energy window) is typically that for the line, as we saw in Sec.~\ref{sec:diffuse-existing}.

The total integrated flux is over the angular coverage of the system of interest and over the energy bin that best constrains the dark matter signal.
We consider sensitivities which can be obtained with two different strategies.
The conservative sensitivity (as in Sec.~\ref{sec:diffuse-existing}) can be made by assuming that diffuse emission may be due solely to dark matter, excluding models that would be expected to produce a number of events in excess ($2 \sigma $) of that measured in any bin.
An optimistic sensitivity can be derived by assuming that the energy spectrum of the astrophysical background exhibits no sharp features and can be determined by fitting to the data.
One then excludes models that would produce an excess of events in any bin even if the astrophysical background in that bin were overestimated by a $2\sigma$ systematic uncertainty.
We assume up to a 15\% systematic error in the determination of the smooth component of the differential flux, which is quoted for EGRET~\cite{Strong:2004de}.
Note that we need not consider statistical fluctuations for the optimistic analysis, since the systematic uncertainties dominate.

The background is estimated from a single power-law fit to the COMPTEL data~\cite{Weidenspointner:1999thesis} over energies $0.8$--$30~\MeV$ and to EGRET data~\cite{Strong:2004de} over energies $30~\MeV$--$10~\GeV$:
\begin{equation}
  \frac{d^2 \Phi}{dE\, d\Omega} = 2.74 \times 10^{-3}
  \left(\frac{E}{\MeV}\right)^{-2.0}~\cm^{-2}\s^{-1}\sr^{-1}\MeV^{-1} \ .
  \label{eq:bkg-fit}
\end{equation}
(Fermi data is not used, since it overlaps and agrees well with EGRET data at energies less than a few hundred MeV.)
The COMPTEL data corresponds to observations over the region $|b|>30^\circ$, while the EGRET data was an average of two regions ($20^\circ<|b|<60^\circ$ and $60^\circ<|b|<90^\circ$), weighted by their relative spans of the sky.
Note that the data from EGRET encompass a slightly larger region compared to COMPTEL, so the fit underestimates the flux at lower energies and overestimates the flux at higher energies.
For $E<150~\MeV$, the EGRET data in the region $20^\circ<|b|<60^\circ$ is only a factor of $1.2$--$1.4$ larger than that in the region $60^\circ<|b|<90^\circ$.
Thus, we use this fit as a rough guide for predicting the diffuse background a future telescope would detect.

The resulting sensitivities are shown in Fig.~\ref{fig:constraints-future} for observing the region $|b|>30^\circ$.
For either the conservative or optimistic analysis, it is important to note that the sensitivity presented here is not determined by the exposure; the maximum differential gamma-ray flux which can be attributed to dark matter is already measured by COMPTEL, EGRET, and Fermi, and we assume statistical uncertainties are subleading.
Instead, the sensitivity is determined by the experiment's energy resolution.
Certainly, more reliable astrophysical modeling in the future will be able to significantly improve sensitivity estimates and constraints on dark matter.
In particular, anisotropies in the extragalactic diffuse background produced from dark matter annihilations are predicted to be different from those produced by standard astrophysical sources~\cite{Ando:2005xg,Ando:2006cr,SiegalGaskins:2009ux,Inoue:2013vza}.
However, in the meantime, we translate our ignorance of the functional form of the diffuse background into the 15\% systematic error previously discussed.

Comparing the results in Fig.~\ref{fig:constraints-future} to those in Fig.~\ref{fig:constraints}, projected sensitivities from GRIPS are better than all present diffuse bounds for $\gamma\gamma$ and $\pi^0\pi^0$.
For $\gamma\gamma$, the monoenergetic photon line lies within the energy range of GRIPS up to $\sqrt{s} \sim 160~\MeV$, and the sensitivities for $\avg{\sigma v}/m_X$ and the lifetime are approximately constant in $\sqrt{s}$.
Similarly, for $\pi^0\pi^0$, the box spectrum of the pion is narrow enough near $\sqrt{s} \sim 270~\MeV$ to mimic a line signal, but contributes less to the sensitivity as the box widens at higher $\sqrt{s}$.
For $\gamma\pi^0$, the pion box provides the dominant contribution to sensitivity until it extends too far above the GRIPS upper threshold, for $\sqrt{s} \gtrsim 160~\MeV$, at which point the photon line gives the better constraint.
GRIPS's sensitivity will exceed present bounds until it suffers a sudden loss of sensitivity for $\sqrt{s} \gtrsim 240~\MeV$, where the photon line is above the GRIPS energy threshold; meanwhile, the photon line is still inside the energy range of EGRET and Fermi.

AdEPT sensitivities are comparable to present diffuse bounds, and it is less sensitive than GRIPS, due to its larger energy resolution.
The exception is in the $\gamma\pi^0$ channel for $\sqrt{s} \gtrsim 240~\MeV$, where the photon lies within the AdEPT energy window but is above the GRIPS upper energy threshold.
In fact, for the mass range plotted, all spectral features from dark matter annihilation/decay are contained in the AdEPT energy window for all three channels.
For $\gamma\pi^0$, the pion box provides the better sensitivity for $\sqrt{s} \lesssim 200~\MeV$, above which the line and box signals overlap.
The energy bin that contains the line also contains part of the box signal, so this bin becomes the more constraining one.
For $\gamma\gamma$, the sensitivities for $\avg{\sigma v}/m_X$ and the lifetime are approximately constant in $\sqrt{s}$, except when $\sqrt{s} \lesssim 10~\MeV$, for which the photon line drops below the AdEPT energy threshold.

ACT sensitivities are the most strict in its relevant mass range for the $\gamma\gamma$ channel, with its conservative sensitivity lying on the CMB limit for annihilation.
Like GRIPS and AdEPT, the sensitivities for $\avg{\sigma v}/m_X$ and the lifetime are approximately constant in $\sqrt{s}$.
For the $\gamma\pi^0$ channel, as with COMPTEL (but unlike GRIPS and AdEPT), sensitivity is weak near the threshold at $\sqrt{s} \sim 135~\MeV$, because the pion box is above ACT's energy range, and the low-energy photon line is surrounded by a large background.
ACT barely does better than EGRET near $\sqrt{s}\sim 145~\MeV$, until the photon leaves the ACT energy range at larger $\sqrt{s}$.
For the $m_X$ range of interest, the box spectrum from $\pi^0\pi^0$ is at much higher energies than the upper energy threshold of ACT; thus, no ACT limits are included for this channel.

The sensitivity of GAMMA-400 is on par with Fermi for the $\gamma\gamma$ channel.
The photon line from $\gamma\pi^0$ is never above GAMMA-400's lower energy threshold in most of the plotted $m_X$ range, so its sensitivity is mainly due to the pion box spectrum.
However, the true upper edge of the pion box only surpasses GAMMA-400's lower energy threshold for $\sqrt{s} \geq 200~\MeV$; for smaller $\sqrt{s}$, GAMMA-400's sensitivity arises from the smearing of the box due to the energy resolution.
Since Fermi has a better energy resolution than GAMMA-400, as quoted in Table~\ref{tab:experiments}, there is more leakage into the energy range of GAMMA-400 for $\sqrt{s}<200~\MeV$, resulting in the greater projected sensitivity for GAMMA-400.
But this result strongly depends on the form of the detector response function and may not hold if the actual response function is significantly different from the one we use [Eq.~\eqref{eq:resp_func}].
The situation for the $\pi^0 \pi^0$ channel is similar, except that there is no plotted region for which the true pion box lies within GAMMA-400's energy window---its sensitivity is due solely to the smearing of the spectrum from the detector energy resolution.

\subsection{Diffuse background in the 0.3--10 MeV range}

We note that the explanation of the origin of the extragalactic background in the $\MeV$ region is incomplete and remains under investigation.
For energies $\lesssim 0.3~\MeV$, the diffuse background can be largely explained by the contribution of active galactic nuclei and Seyfert galaxies~\cite{Madau:1994zz,Ueda:2003yx,Inoue:2007tn,Inoue:2013vza}, while for energies $\gtrsim 10~\MeV$ it is believed that blazars~\cite{Zdziarski:1996eu,Sreekumar:1997un,Ajello:2009ip} and radio and star-forming galaxies~\cite{Ajello:2015mfa,DiMauro:2015tfa} can provide an adequate explanation for the observed background.
There is also a small but insufficient contribution from Type Ia supernovae~\cite{Horiuchi:2010kq,Ruiz-Lapuente:2015yua} near $1~\MeV$.
As a result, there is no clear explanation for the observed (sharply falling) spectrum in the intermediate range $0.3$--$10~\MeV$.

The class of dark matter models in Sec.~\ref{sec:model} can contribute to the observed photon flux in this energy range.
However, the spectral signatures of redshifted gamma-ray lines and boxes alone are not suitable to explain the sharply decreasing intensity of the observed background.
It would be interesting to consider whether these distinctive features from dark matter annihilation/decay, in combination with contributions from astrophysical sources, could adequately explain the observed photon spectrum in this energy range.
A more detailed analysis is beyond the scope of this work.

\section{Indirect Detection: Dwarf Spheroidal Galaxies}
\label{sec:dwarf}

The Milky Way satellite dwarf spheroidal galaxies are dark matter dominated and thus very faint, so we do not expect any significant source of photons from dwarfs due to standard astrophysical processes, making dwarfs a very clean set of systems to study.
Although the expected signal is low, the dwarfs lie away from the Galactic plane, so the astrophysical backgrounds are much smaller than they would be otherwise.
Importantly, one can estimate the astrophysical background from data by looking slightly off axis.
Table~\ref{tab:dwarfs} shows the integrated $J$-factors for 20 dwarfs, from Ref.~\cite{GeringerSameth:2014yza}.
The total flux from a dwarf is contained within an angle $\theta_\textrm{max}$, set by the outermost observed star in the halo, from the center of the dwarf.
We consider regions $\theta > \theta_\textrm{max}$ to give zero contribution to the density profile, resulting in a conservative value for the integrated $J$-factor.
In determining the dark matter flux measured by a detector, the angular integration of the $J$-factor extends over the region defined by the halo or by the detector point spread function (PSF) 68\% containment radius, whichever of the two is larger.
We choose Draco as our case study for two main reasons: it has large integrated $J$-factors for both annihilation and decay, and independent calculations of $J$ agree well with one another~\cite{Ackermann:2011wa,Martinez:2013els,GeringerSameth:2014yza}.
Comparison with other dwarfs is straightforward using Table~\ref{tab:dwarfs}; in particular, for a given observed flux, the corresponding dark matter annihilation cross section or decay width scales inversely with the $J$-factor in Eq.~\eqref{eq:diff2flux}.
Performing a stacked analysis from the observation of many dwarfs will further improve sensitivity, though such a detailed study is beyond the scope of this work.

\begin{table}[t]
  \centering
  \begin{tabular}{lccccc}
    Dwarf & longitude $l$ & latitude $b$ & $\theta_\textrm{max}$
    & $\log_{10}\mathcal{J}_\textrm{ann}$ & $\log_{10}\mathcal{J}_\textrm{dec}$ \\
    & [deg] & [deg] & [deg] & $[\GeV^2 \cm^{-5}]$ & $[\GeV \cm^{-2}]$ \\
    \hline
    \hline
    Bootes I & 358.08 & 69.62 & 0.47
    & $18.24^{+0.40}_{-0.37}$ & $17.90^{+0.23}_{-0.26}$ \\
    Carina & 260.11 & -22.22 & 1.26
    & $17.92^{+0.19}_{-0.11}$ & $18.15^{+0.34}_{-0.25}$ \\
    Coma Berenices & 241.89 & 83.61 & 0.31
    & $19.02^{+0.37}_{-0.41}$ & $17.96^{+0.20}_{-0.25}$ \\
    Canes Venatici I & 74.31 & 79.82 & 0.53
    & $17.44^{+0.37}_{-0.28}$ & $17.57^{+0.37}_{-0.73}$ \\
    Canes Venatici II & 113.58 & 82.70 & 0.13
    & $17.65^{+0.45}_{-0.43}$ & $16.97^{+0.24}_{-0.23}$ \\
    Draco & 86.37 & 34.72 & 1.30
    & $19.05^{+0.22}_{-0.21}$ & $18.97^{+0.17}_{-0.24}$ \\
    Fornax & 237.10 & -65.65 & 2.61
    & $17.84^{+0.11}_{-0.06}$ & $17.99^{+0.11}_{-0.08}$ \\
    Hercules & 28.73 & 36.87 & 0.28
    & $16.86^{+0.74}_{-0.68}$ & $16.66^{+0.42}_{-0.40}$ \\
    Leo I & 225.99 & 49.11 & 0.45
    & $17.84^{+0.20}_{-0.16}$ & $17.91^{+0.15}_{-0.20}$ \\
    Leo II & 220.17 & 67.23 & 0.23
    & $17.97^{+0.20}_{-0.18}$ & $17.24^{+0.35}_{-0.48}$ \\
    Leo IV & 265.44 & 56.51 & 0.16
    & $16.32^{+1.06}_{-1.70}$ & $16.12^{+0.71}_{-1.13}$ \\
    Leo V & 261.86 & 58.54 & 0.07
    & $16.37^{+0.94}_{-0.87}$ & $15.87^{+0.46}_{-0.47}$ \\
    Leo T & 214.85 & 43.66 & 0.08
    & $17.11^{+0.43}_{-0.39}$ & $16.48^{+0.22}_{-0.25}$ \\
    Sculptor & 287.54 & -83.16 & 1.94
    & $18.57^{+0.07}_{-0.05}$ & $18.47^{+0.16}_{-0.14}$ \\
    Segue 1 & 220.48 & 50.43 & 0.35
    & $19.36^{+0.32}_{-0.35}$ & $17.99^{+0.20}_{-0.31}$ \\
    Segue 2 & 149.43 & -38.14 & 0.19
    & $16.21^{+1.06}_{-0.98}$ & $15.89^{+0.56}_{-0.37}$ \\
    Sextans & 243.50 & 42.27 & 1.70
    & $17.92^{+0.35}_{-0.29}$ & $18.56^{+0.25}_{-0.73}$ \\
    Ursa Major I & 159.43 & 54.41 & 0.43
    & $17.87^{+0.56}_{-0.33}$ & $17.61^{+0.20}_{-0.38}$ \\
    Ursa Major II & 152.46 & 37.44 & 0.53
    & $19.42^{+0.44}_{-0.42}$ & $18.39^{+0.25}_{-0.27}$ \\
    Ursa Minor & 104.97 & 44.80 & 1.37
    & $18.95^{+0.26}_{-0.18}$ & $18.13^{+0.26}_{-0.18}$
  \end{tabular}
  \caption{Integrated $J$-factors for dwarf galaxies from Ref.~\cite{GeringerSameth:2014yza}.
    The integration is centered on the dwarf and extends out to an angle of $\theta_\textrm{max}$, the point at which the halo density profile is truncated.
    The $1\sigma$ errors are due to the uncertainty in the precise shape of the profile.
    The location of the dwarfs are given by their Galactic coordinates $(l,b)$, from the NASA/IPAC Extragalactic Database (\url{http://ned.ipac.caltech.edu/}).}
  \label{tab:dwarfs}
\end{table}

There are recent constraints on dark matter annihilation from searches in dwarf galaxies using Fermi data.
Reference~\cite{GeringerSameth:2014qqa} performed an analysis for annihilation to $\gamma\gamma$, and the limit they derived is comparable to CMB limits (see their Fig.~11), but the analysis was for $m_X > 1~\GeV$ and thus it is not included here.

\subsection{Limits from future experiments}

For analyzing Draco, the relevant energy bins are all the ones encompassing the dark matter signal, whether they are from the monoenergetic photon or from the box spectrum from pion decay, since there are relatively low statistics expected from dwarf observations.
We again use an optimum bin for the photon line, as discussed in Sec.~\ref{sec:diffuse-future}, and we define a single large ``bin'' for the box that spans energies from $(1-\epsilon)E_-$ to $(1+\epsilon)E_+$, unless detector energy thresholds truncate the box spectrum.
To find the total integrated flux, the angular integration is over the PSF of the detector.
The integrated $J$-factor for a PSF $\gtrsim 1^\circ$ is very close to simply using $\theta_\textrm{max}$ instead.
Thus, despite Draco having one of the largest extended profile emissions~\cite{GeringerSameth:2014yza}, we may still treat it as a point source.

\begin{figure}[t]
  \includegraphics[width=0.95\textwidth]{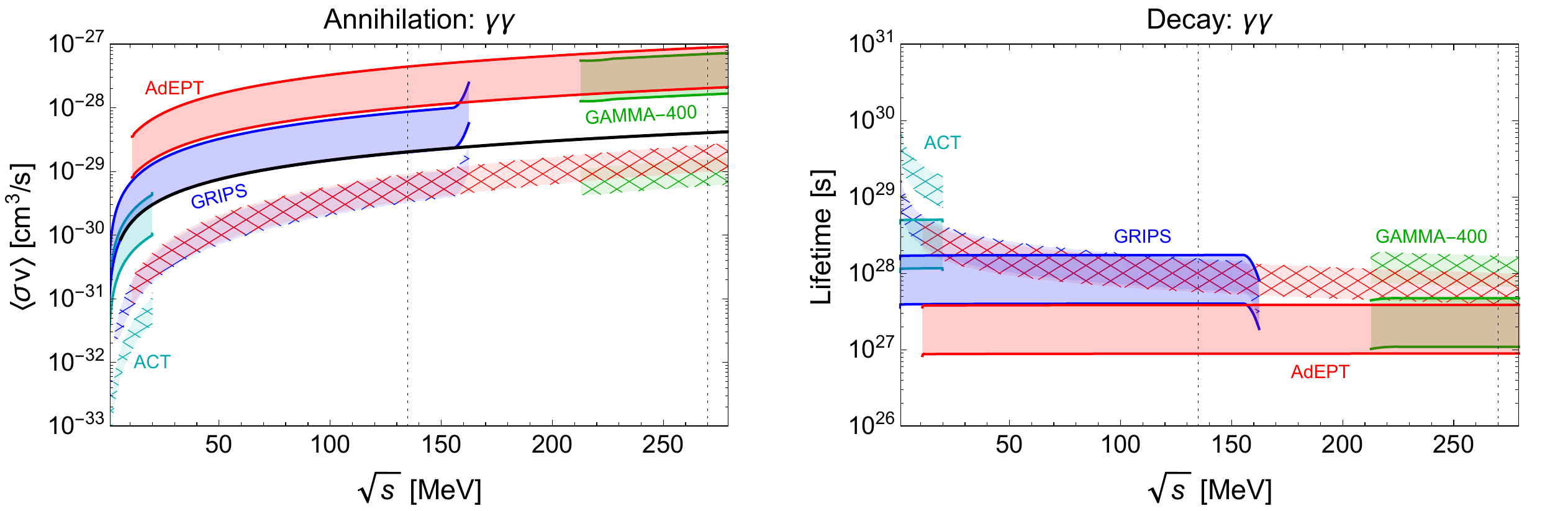} \\
  \includegraphics[width=0.95\textwidth]{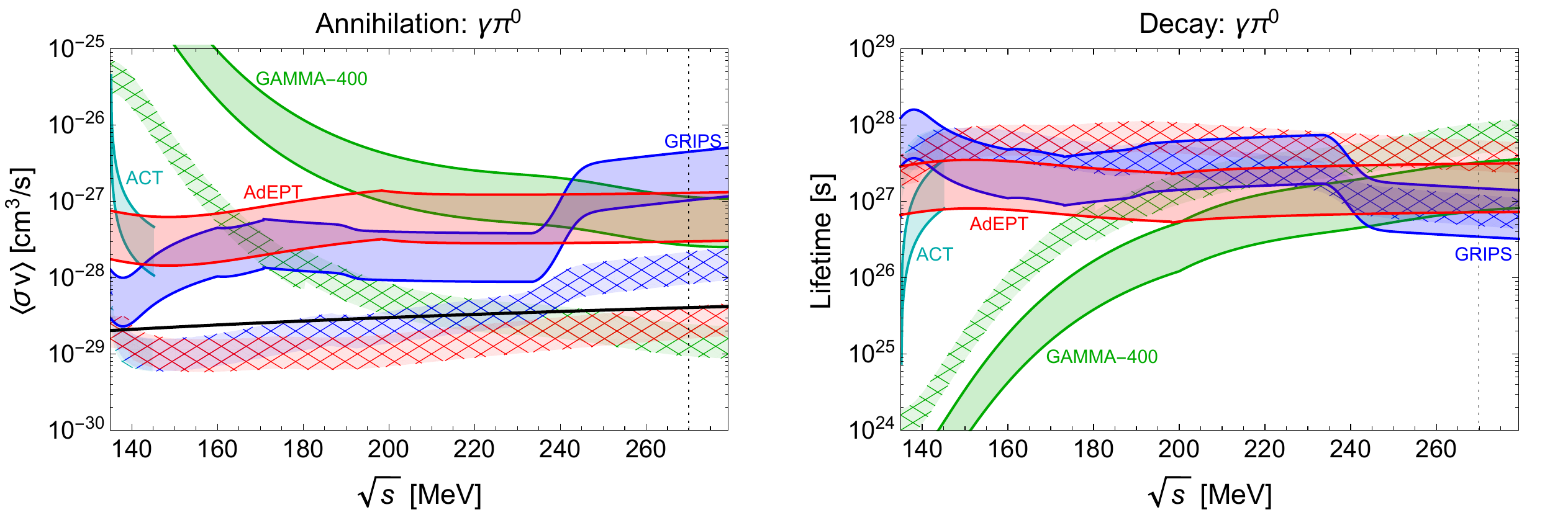} \\
  \includegraphics[width=0.95\textwidth]{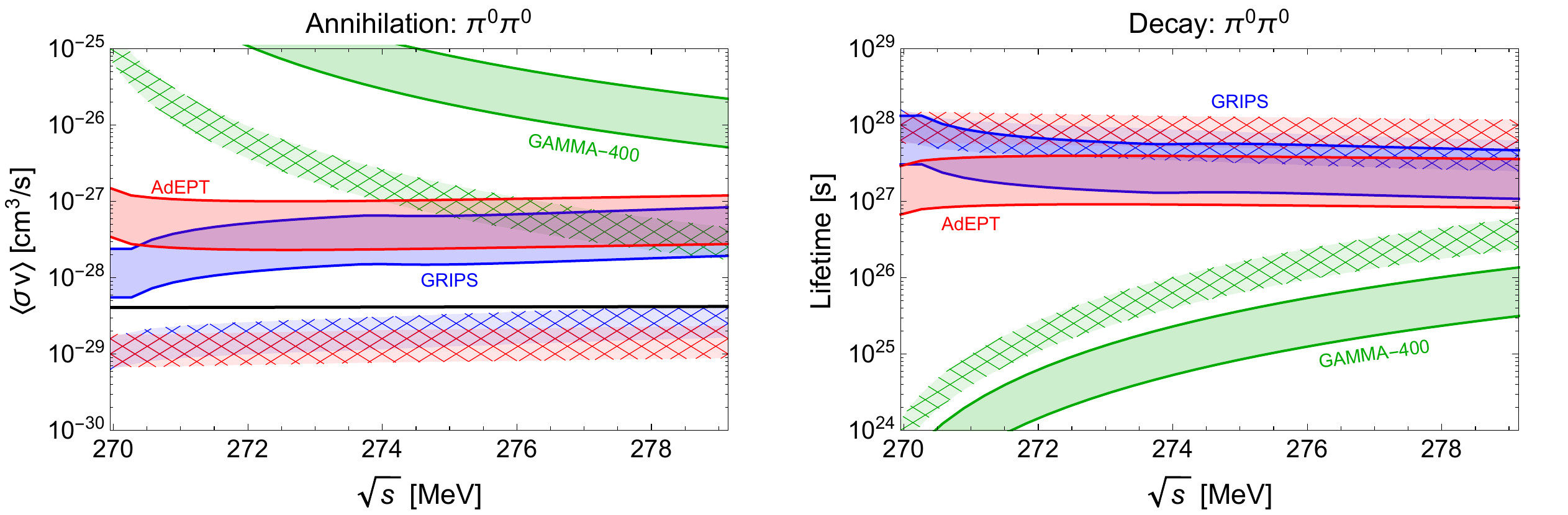}
  \caption{Projected sensitivities on dark matter annihilation (left) and decay (right) for channels $\gamma\gamma$ (top), $\gamma\pi^0$ (middle), and $\pi^0 \pi^0$ (bottom).
  Planck (black) bounds are shown from Fig.~\ref{fig:constraints}.
  The cyan, blue, red, and green shaded regions show the areas between the conservative and optimistic sensitivities for the diffuse background for ACT, GRIPS, AdEPT, and GAMMA-400, respectively.
  The hatched regions show the $1\sigma$ uncertainty bounds for Draco after five years of observation.
  Vertical dotted lines show the kinematic thresholds at $m_{\pi^0}$ and $2m_{\pi^0}$ (plotting begins at $m_{\pi^0}$ for the $\gamma\pi^0$ channel and at $2m_{\pi^0}$ for the $\pi^0 \pi^0$ channel), and $\sqrt{s}$ is plotted up to $m_{\pi^\pm}$ for all three channels.
  Note that the vertical axes cover a different range for $\gamma\gamma$ than for the other two channels.}
  \label{fig:constraints-future}
\end{figure}

We assume that the only relevant background is the diffuse emission, and unresolved point sources contribute negligibly.
Any diffuse contribution from dark matter is already incorporated in the data, so the fit \eqref{eq:bkg-fit} represents the total background for observing dwarfs.
If the expected number of events for a model in relevant energy bins (including signal and background) is $\mu$, then the probability of the model not yielding an excess of events above background is
\begin{equation}
  P(n\leq n_\textrm{obs}) = \sum_{n=0}^{n_\textrm{obs}} \frac{1}{n!} e^{-\mu} \mu^n \ ,
\end{equation}
where $n_\textrm{obs}$ is the number of observed events, estimated by Eq.~\eqref{eq:bkg-fit}.
We then find the 95\% exclusion limits for the annihilation cross section or decay rate for five years of running.
The limits are plotted as hatched regions in Fig.~\ref{fig:constraints-future}.
The band thickness represents the $1\sigma$ systematic uncertainty for the integrated $J$-factor in Table~\ref{tab:dwarfs}.

As expected, dwarfs are the more advantageous target if dark matter annihilates, since the annihilation rate scales as the square of the dark matter density; the large density of dark matter within a dwarf then provides a large enhancement to the photon production rate.
In fact, ACT, GRIPS, and AdEPT sensitivities of the annihilation cross section from Draco are able to reach the bound from Planck's CMB observations, except when GRIPS loses sensitivity near $\sqrt{s} \sim 240~\MeV$ in the $\gamma\pi^0$ channel.
Additionally, AdEPT has a comparable or better sensitivity than GRIPS in all three channels, despite its sensitivity being worse for the diffuse background.
This result reflects the fact that counting statistics are the determining factors for the dwarf sensitivities, giving dwarfs the advantage over the systematics-dominated diffuse background.
AdEPT is helped by its larger $A_\textrm{eff}$ and by its larger energy window, compensating for its poorer energy resolution.

On the other hand, the dark matter decay rate only scales as one power of the density, so the sensitivities derived from a dwarf versus diffuse analysis should not differ as greatly as they do for annihilation.
Indeed, the diffuse and dwarf decay sensitivities are comparable for GRIPS, and the dwarf sensitivity is only slightly better for ACT, AdEPT, and GAMMA-400.

\section{A Comparison to Collider-Based Bounds}
\label{sec:collider}

The interactions of low-mass dark matter with first-generation quarks can be tightly constrained by ``monoanything'' searches~\cite{Feng:2005gj,Goodman:2010yf,Bai:2010hh,Goodman:2010ku,Cheung:2012gi,Zhou:2013fla} at the LHC~\cite{Aad:2012ky,Chatrchyan:2012me,ATLAS-CONF-2012-147,Aad:2013oja,CMS-PAS-EXO-12-048,Aad:2014vka}.
However, the sensitivity of these searches depends on the details of the dark-matter--quark effective interaction; it is thus difficult to compare such constraints to the results of the analysis presented here, which only depends on the annihilation/decay branching fraction to each channel.
However, one can gain some insight into the relative sensitivity of indirect and collider search strategies for a particular model, namely, fermionic dark matter which couples to first-generation quarks via an effective contact operator, given by ${\cal O} = (1/M^2)(\bar X \gamma^\mu X)(\bar q \gamma_\mu q)$.

If the effective contact operator approximation is valid, then data from ATLAS require $M \gtrsim 800~\GeV$ for low-mass dark matter~\cite{ATLAS-CONF-2012-147}.
This coupling permits dark matter annihilation to the $\gamma \pi^0$ final state via the effective operator given in Eq.~\eqref{eq:VV_eff_op}.
If we require that for $\sqrt{s} = 2m_{\pi^0}$ this cross section obey $\langle \sigma_A v \rangle \sim 10^{-3}~\pb$ (roughly the optimistic sensitivity of GAMMA-400 searching for a signal from Draco), then one finds $M \sim 10~\GeV$.
As such, if the contact operator approximation is valid for energies ${\cal O}(10^4~\GeV)$, then the sensitivity of the LHC may be expected to far exceed any search of ${\cal O}(1-100)~\MeV$ gamma rays.
But for such light dark matter, it is easily possible that the scale of the mediating particles is very light compared to the energy scale of the LHC.
If $M_*$ is the mass of a single mediating particle exchanged in the $s$ channel, then one roughly finds $M \sim M_* / \sqrt{g_X g_q}$, where $g_X$ and $g_q$ are the coupling of the mediator to dark matter and first-generation quarks, respectively.
As a result, the mediator $M_*$ can be ${\cal O}(100~\MeV)$, provided the coupling $g_q$ is sufficiently small.
In that case, standard ``monoanything'' LHC searches would be unconstraining, and it is not clear if there will be any signal of new physics at the LHC.

In particular, if the dark matter annihilation process $\bar X X \rightarrow \gamma \pi^0$ proceeds in the $s$ channel with a cross section $\sim 10^{-3}~\pb$, then the mediator could be as light as $\sim 300~\MeV$ with $g_X \sim 1$, provided $g_q \sim 10^{-3}$.
It would be very challenging for the LHC to find evidence for such a weakly coupled light mediator above the background of electroweak interactions, which can produce missing transverse momentum via neutrinos.

\section{Conclusions}
\label{sec:conclusions}

In this work, we have considered the indirect-detection signals which may be seen by gamma-ray telescopes, sensitive to the range ${\cal O}(1-100)~\MeV$ for models where light dark matter couples to first-generation quarks.
For dark matter with mass less than $m_{\pi^\pm}$ ($2m_{\pi^\pm}$), dark matter annihilation (decay) is constrained by kinematics to yield very simple final states involving pions and photons, resulting in a very distinctive photon signature.

For the mass range we consider, dark matter $s$-wave annihilation with $\langle \sigma v \rangle \sim 1~\pb$ is already ruled out by constraints from Planck, as well as by constraints from COMPTEL, EGRET, and Fermi.
But if the process by which the dark matter density is generated is nonthermal, then the dark matter annihilation cross section may be much smaller and may be probed by future experiments, such as ACT, GRIPS, and GAMMA-400.
The most promising strategy is observing dwarf spheroidal galaxies, due to the possibility of data-driven background subtraction.
A search for diffuse emission of photons from dark matter $s$-wave annihilation may also yield new constraints which are better than those currently available, but the sensitivity of this strategy is limited by one's ability to accurately model the astrophysical background.
However, future telescopes aiming for better sensitivities are expected to---by construction---more tightly constrain dark matter $p$-wave annihilation, for which current bounds are set by existing diffuse searches.

If dark matter decays, then one finds, as expected, that dwarf spheroidal galaxies are not as attractive of a target, because the decay rate only scales as the density.
In this case, bounds from diffuse emission searches may be comparable to those from dwarf spheroidal searches, provided one can make some reasonable estimate for the contribution to the photon flux arising from astrophysical backgrounds.

For annihilation (decay) to the $\gamma \pi^0$ or $\pi^0 \pi^0$ channels, upcoming telescopes could be sensitive to models with $\langle \sigma v \rangle \sim 10^{-4}~\pb$ ($\tau \sim 10^{28}~\s$).
Sensitivity in the $\gamma \gamma$ channel could be up to an order of magnitude better.
These represent vast improvements (up to a few orders of magnitude) over current sensitivity from gamma-ray telescopes and from cosmological observations.
For dark matter models with $m_X \leq 2m_{\pi^\pm}$, current direct-detection experiments have limited sensitivity, while the sensitivity of collider searches are highly dependent on the details of the coupling of dark matter to the SM.
On the other hand, indirect searches for gamma rays arising from dark matter annihilation or decay can be ideal probes of this class of models, which generically provide striking photon spectra.
This result motivates the development of future telescopes that will fill the ``MeV gap'' in sensitivity and will be able to robustly test these models.

\section*{Acknowledgments}

We are grateful to John Beacom, Danny Marfatia, Arvind Rajaraman, Patrick Stengel, Louie Strigari, and Xerxes Tata for useful discussions.
This research is funded in part by NSF CAREER grant PHY-1250573.
J.~K. is grateful to Indiana University, where part of this research was performed, for their hospitality.
This research has made use of the NASA/IPAC Extragalactic Database (NED), which is operated by the Jet Propulsion Laboratory, California Institute of Technology, under contract with the National Aeronautics and Space Administration.

\bibliography{physics-refs}
\bibliographystyle{notitle}
\end{document}